# Interaction between the small gas molecules and the defective WSe$_2$ monolayer

Dongwei Ma[1,*], Chaozheng He[2,*], Benyuan Ma[2], Zhiwen Lu[2], Yanan Tang[3,*], Zhansheng Lu[4] and Zongxian Yang[4]

[1]*School of Physics, Anyang Normal University, Anyang 455000, China*

[2]*Physics and Electronic Engineering College, Nanyang Normal University, Nanyang 473061, China*

[3]*College of Physics and Electronic Engineering, Zhengzhou Normal University, Zhengzhou, 450044, China*

[4]*College of Physics and Materials Science, Henan Normal University, Xinxiang 453007, China*

**Abstract**

In this study, the interaction between the gas molecules, including H$_2$O, N$_2$, CO, NO, NO$_2$ and N$_2$O, with the WSe$_2$ monolayer containing a Se vacancy (denoted as V$_{Se}$) is theoretically studied. It is found that H$_2$O and N$_2$ molecules are highly prone to be physisorbed on the V$_{Se}$ surface. The presence of the Se vacancy should significantly enhance the sensing ability of the WSe$_2$ monolayer toward the H$_2$O and N$_2$ molecules. In contrast, CO and NO molecules highly prefer to be molecularly chemisorbed on the V$_{Se}$ surface with the non-oxygen atom occupying the Se site. Further, the exposed O atoms of the molecularly chemisorbed CO or NO can be reacted with the additional CO or NO molecule, to produce the C-doped or N-doped WSe$_2$ monolayer. Our electronic structure calculations show that the WSe$_2$ monolayers are p-doped by the CO and NO molecules, as well as the C and N atoms. The calculated energies suggest that the filling of the CO or NO molecule and the removal of the exposed O atom by the additional CO or NO molecule are both energetically and dynamically favorable. However, only the NO molecule and N atom doped WSe$_2$ monolayers exhibit significantly improved electronic structures compared with the defective WSe$_2$ monolayer. The NO$_2$ and N$_2$O molecules will dissociate directly to form a O-doped WSe$_2$ with a physisorbed NO or

*Corresponding author. E-mail: dwmachina@126.com (Dongwei Ma).
*Corresponding author. E-mail: hecz2013@nynu.edu.cn (Chaozheng He).
*Corresponding author. E-mail: yntang2010@163.com (Yanan Tang).

$N_2$. After desorbing the physisorbed NO or $N_2$, the O-doped $WSe_2$ monolayer can be obtained, for which the defect levels due to the Se vacancy can be completely removed. The calculated energies suggest that although the dissociation processes for $NO_2$ and $N_2O$ molecules are highly endothermic, the $N_2O$ dissociation may need to operate at an elevated temperature compared with the room temperature, due to its large energy barrier of ~ 1 eV.

## 1. Introduction

Investigations on the excellent properties of graphene have stimulated great interest in exploring novel properties and applications of other two-dimensional (2D) layered materials. Graphene is known having an intrinsic zero bandgap, which has hindered significantly its application in logical electronic devices.[1] The layered transition metal dichalcogenides (TMDs) have gained numerous attention in recent years due to its exciting physical and chemical properties.[2-4] Especially, group 6 TMDs, such as $WSe_2$ and $MoS_2$, present remarkable features, such as indirect-to-direct bandgap crossover,[5, 6] layer-dependent tunable bandgap,[5, 7] valley-selective optical excitation,[8, 9] and topological superconductivity[10]. These unique properties make TMDs great promising in the fields of field-effect transistors,[7, 11] flexible electronics,[6, 12] optoelectronics,[13, 14] valleytronics,[15, 16] and electrocatalysts and photocatalysts[17].

$WSe_2$ monolayer has a direct bandgap of ~ 1.6 eV, which is smaller than that of $MoS_2$ monolayer (~ 1.8 eV).[18] $MoS_2$ mostly exhibits unipolar n-type transport characteristics resulting from the pronounced Fermi level pinning effect at the interface between metal and $MoS_2$.[19] On the contrary, $WSe_2$ can achieve electron-dominated, ambipolar, and hole-dominated transport behaviors by engineering the number of layers[20] and simply choosing a suitable contact metal[21, 22]. Field effect transistor experiments have shown respectable mobility values of $WSe_2$ for electrons (110 cm$^2$V$^-$

$^1S^{-1}$)[23] and holes (240 cm$^2$V$^{-1}$S$^{-1}$)[24]. The unique electronic properties, and other excellent attributes have led to a significant increase of the research interest in WSe$_2$ recently.[9, 25, 26]

Due to their atomic thickness, 2D layered semiconductors have inherent large surface-to-volume ratio, which is highly favorable for the gas molecule adsorption and surface-related applications, such as catalysis, sensing, and doping.[17, 27, 28] It is noted that defects with dangling or strained bonds often exhibit much enhanced chemical activity towards the adsorbates compared with the perfect lattice sites.[29-32] For TMD 2D semiconductors, there are have lots of studies investigating the interaction between the common gas molecules (such as H$_2$O, O$_2$, N$_2$, CO, NO, NO$_2$) and the perfect or defective MoS$_2$ monolayer, both experimentally[33-37] and theoretically[29, 38-44].

It is known that chalcogen vacancies are the most commonly occurring point defects in most TMDs. These chalcogen vacancies can create deep acceptor states that act as efficient electron traps and limit the electron mobility.[45-49] Previous reports have shown that the interaction of common gas molecules with the chalocgen vacancies can improve the electronic properties of the as-prepared MoS$_2$. For example, the strong photoluminescence enhancement of the MoS$_2$ monolayer has been observed through defect engineering and oxygen bonding, which is probably formed by the adsorption of O$_2$ molecules at the S vacancies of MoS$_2$.[29, 50, 51] The S vacancies can also be repaired by CO, NO, or NO$_2$ molecules, according to the first-principles calculations based on density functional theory (DFT).[42] On the other hand, the S vacancies may have a beneficial effect for other applications. For example, DFT calculations predicted that the S vacancies can improve the sensing of the common gas molecules.[43, 44] A combination of experimental and theoretical study indicates that the S vacancy allow us to achieve the highest hydrogen evolution reaction activity among all the MoS$_2$ catalyst.[37]

For WSe$_2$, to our knowledge, there are few studies on the interaction between the common gas molecules with its surfaces. WSe$_2$ monolayer has been p-doped using NO$_2$, as revealed by the field-effect transistor experiment.[24] Ovcharenko et al. found a weak adsorption of $NO_2$ on the $WSe_2$ surface and a slight change of the electronic structure of the material by DFT and photoelectron spectroscopy studies.[52] Zhao et al. explored the p-doping of $WSe_2$ via chemisorption of $NO_2$ at a sample temperature of 150 °C. The DFT calculations and spectroscopy results has identified the chemisorbed NO as the dominant doping species at the Se vacancies sites with the N atom occupying the vacancy.[53] Theoretically, Liu et al. studied the transition from chemisorption to physisorption (PS) states of $O_2$ on the $WSe_2$ monolayer with a single Se vacancy (denoted as $V_{Se}$),[29] while Zhao et al. studied the dissociation of the chemisorbed $O_2$ on the $V_{Se}$ surface.[54] The adsorption of several common gas molecules has been studied by Wang et al.[55] on the pristine $WSe_2$ surface (p-$WSe_2$).

In this study, we investigated the effect of Se vacancies on the interaction between the gas molecules, including $H_2O$, $N_2$, CO, NO, $NO_2$ and $N_2O$, with the $WSe_2$ monolayer surfaces. Our study reveals a markedly different interaction behavior of the various gas molecules on the VSe surfaces. For example, the $H_2O$ and $N_2$ molecules are highly prone to be physisorbed on the $V_{Se}$ surface. In contrast, CO and NO molecules highly prefer to be molecularly chemisorbed on the $V_{Se}$ surface with the non-oxygen atom occupying the Se site. Further, the exposed O atoms of the molecularly chemisorbed CO or NO can be reacted with the additional CO or NO molecule, to produce the C-doped or N-doped $WSe_2$ monolayer. The NO2 or N2O molecules will dissociate directly to form a O-doped WSe2 with a physisorbed NO or $N_2$. In the following, the energies, geometric structures and electronic structure of the different adsorption states for various molecules will be presented and discussed. It is hoped our study will give novel insight into the doping of $WSe_2$ via the interaction between common gas molecules and the Se vacancy, and also reveal the possible application of the defective $WSe_2$ in the gas sensing field.

## 2. Computational details

First-principles calculations have been performed using the projected augmented wave formalism of DFT, as implemented in the Vienna Ab-initio Simulation Package (VASP).[56-58] To correctly describe the effect of the van der Waals (vdW) interaction between the adsorbed molecule and the $WSe_2$ monolayer, the optB86b-vdW exchange functional is adopted for the vdW correction,[59, 60] which has been used for the $MoS_2$ systems recently.[61-63] The cutoff energy for the plane-wave basis set is taken as 450 eV. With these settings, the calculated lattice parameter of the $WSe_2$ monolayer is 3.295 Å, which is in good agreement with the previous theoretical values of 3.29 Å[64] and 3.298[65]. Band structure calculations show that the pristine $WSe_2$ monolayer is an direct bandgap semiconductor with the bandgap of 1.62 eV, which is also in agreement with previous theoretical (1.60 eV)[64] and experimental (1.65 eV) results.[66]

For the $V_{Se}$ system, a (4×4) supercell is used with a single Se atom removed. The distance between the defective $WSe_2$ monolayer and its neighboring image is larger than 18 Å, which is sufficiently large to avoid the interaction between them. The Monkhorst-Pack grid of (8×8×1) is adopted for the calculations of densities of states (DOS) and (3×3×1) for others.[67] The total energy convergence is considered to be achieved until two iterated steps with energy difference less than $10^{-5}$ eV. Structure optimizations were performed until the Hellmann-Feynman force on each atom is less than 0.02 eV/Å. During the optimization, all the internal coordinates are allowed to relax with a fixed lattice constant. Furthermore, the climbing image nudged elastic band method[68] is used to find the minimum-energy path for transition between different stable states. The spring constants between adjacent images are set to -5.0 eV/Å$^2$.

## 3. Results and discussion

### 3.1 Properties of $V_{Se}$

The formation, and the geometrical and electronic properties of the $V_{Se}$ system are first presented and discussed. As stated above, the single Se vacancy is produced by removing one Se atom from the (4×4) supercell of the pristine $WSe_2$ monolayer. The vacancy formation energy is studied and defined as: $E_{form}= E_{tot}(V_{Se}) − E_{tot}(pristine) + \mu_{Se}$, where $E_{tot}(V_{Se})$ and $E_{tot}(pristine)$ are the total energies of $V_{Se}$ and the pristine $WSe_2$ monolayer, respectively. $\mu_{Se}$ is the chemical potential of the removed Se. The Se rich and W rich conditions have been considered for calculation of the formation energy of the Se vacancy. At the Se-rich limit, $\mu_{Se}$ is equal to the total energy per Se atom in its bulk reference phase ((i. e. the Se bulk having hexagonal crystal lattice with space group no. 152)[69]. At the W rich-limit, $\mu_W$ is equal to the total energy per W atom in its body center cubic bulk phase, while $\mu_{Se}$ has been calculated by $\mu_{WSe2} = \mu_W + 2\mu_{Se}$ ($\mu_{WSe2}$ equal to total energy per formula unit of the pristine monolayer WSe$_2$). With such a definition, defect with low formation energies will occur in high concentrations.[70] The calculated vacancy formation energy at the W-rich limit is 2.04 eV, which is close to the result (~ 1.8 eV) in Ref.[70]. For the Se-rich limit, the formation energy for the Se vacancy is 2.09 eV. This value is much smaller than the result (~ 2.7 eV) of Ref.[70], which may result from different Se bulk reference phase has been used. Nevertheless, the calculated formation energies for the Se vacancy are comparable to the S vacancy in the MoS$_2$ monolayer. For example, Liu et al. predicted that the formation energies of a S vacancy at the S-poor and S-rich limits are 2.35 and 0.95 eV, respectively.[71] The formation energy by Haldar et al. lies between about 1.3 and 2.6 eV depending on the chemical potential of the element S.[70] This suggests that the equilibrium concentrations of the Se vacancy in the WSe$_2$ monolayer should be similar to those of the S vacancy in the monolayer MoS$_2$. Consequently, the presence of the Se vacancy will also play an important role in the electronic and optical properties of 2D WSe$_2$ semiconductors, similar to the role of S vacancies in the monolayer MoS$_2$ electronics,[72] which has been demonstrated in the experimental studies.[53, 73-75]

The optimized structures for the V$_{Se}$ systems are shown in Fig. 1. It can be seen that there is no obvious reconstruction around the Se vacancy, which is similar to the S vacancy in the MoS$_2$[48] while quite distinct from the graphene with a single C vacancy[76]. The most obvious change of the geometrical structure due to the Se vacancy may be that the W atoms surrounding the vacancy move toward the vacancy, such that the W-W distance is reduced by 0.23 Å compared with the pristine WSe$_2$ monolayer. However, the bond lengths between the under-coordinated W atoms and its bonded Se atoms is only reduced by 0.02 Å and they differ only by $10^{-5}$ Å. Therefore, the 3-fold symmetry has been maintained for the V$_{Se}$ system.

For the electronic structure, the calculated band structures of V$_{Se}$ are shown in Fig. 1(b). Compared with the pristine WSe$_2$ monolayer (not shown), the bandgap exhibits a negligible change into 1.61 eV, and the top valence band (VB) and the bottom conduction band (CM) are only slightly disturbed. The most obvious effect due to the

Se vacancy is that there are flat defect states appearing ~ 0.4 eV below the bottom of the CM, which is typical for most TMDs. The total DOS for $V_{Se}$ and the local DOS (LDOS) projected on the W atoms surrounding the Se vacancy are shown in Fig. 1(c). The LDOS indicates that the deep acceptor states in the bandgap originate from the W dangling bonds ($5d$ states). As stated in the previous studies, such levels can reduce the carrier mobility, by trapping or scattering other charge carriers, and lead to a decrease of the quantum efficiency of the material.[29, 49, 74] On the other hand, the dangling bonds often exhibit high chemical activity towards the external adsorbates, compared with the perfect lattice sites,[29-32] which may provide new opportunities for doping of materials, and their applications in surface-related applications, such as sensing and catalysis.

**3.2 Physisorption of $H_2O$ and $N_2$ molecules**

To study the adsorption stability of the considered molecules on the defective $WSe_2$ monolayer, the adsorption energy is defined as: $E_{ad} = E_{vac} + E_{mol} - E_{vac-mol}$, where $E_{vac}$ and $E_{vac-mol}$ are the total energies of the defective $WSe_2$ monolayers without and with the adsorbed molecules, respectively, and $E_{mol}$ is the total energy of the free molecule. With this definition, a positive value of the adsorption energy indicates that the process is exothermic and energetically favorable. For the adsorption of gas molecules, three adsorption states have been considered for each molecule, i.e., PS, molecular chemisorption (MC), and dissociative chemisorption (DC). The calculated adsorption energies for all the considered molecules at the available states are shown in Table 1.

For $H_2O$ molecules, it is found that the molecule can only be physisorbed above the surface of $V_{Se}$. If the molecule is put into the vacancy in the initial configuration, it will run away from the vacancy site after structural optimization. The optimized structures for $H_2O$ adsorption on $V_{Se}$ are shown in Fig. 2(a). The $H_2O$ molecule is 0.88 Å away from the surface and adopts a vertical configuration, where one O-H bond is almost perpendicular to the $WSe_2$ basal plane with the H atom pointing to the Se vacancy site. Such configuration is different from the cases of $H_2O$ molecule adsorption on the perfect $MoS_2$ and $WSe_2$ monolayers (as presented in Fig. S1 (a) of the Electronic

Supporting Information (ESI)), where the axis that connects two H atoms is parallel to the basal planes of the monolayer sheets. The H-O-H bond angle is reduced by 0.2º. The bond length of the O-H bond almost perpendicular to the $WSe_2$ surface is elongated by 0.01 Å, while another O-H bond is almost unchanged. In addition, the calculated adsorption energy for $H_2O$ molecules is 0.30 eV. These results suggest that the $H_2O$ molecule is physisorbed on the surface of $V_{Se}$. For the $N_2$ molecule, except the PS state, the MC state can be also obtained. However, the MC state is endothermic by 0.08 eV and is less stable than the PS state by 0.25 eV. Thus, we only focus on the PS state. The optimized structures for $N_2$ adsorption on $V_{Se}$ are shown in Fig. 2(b). The distance between the $N_2$ molecule and the surface of $V_{Se}$ is 2.39 Å, much larger than that for $H_2O$. Accordingly, the $N_2$ molecule (0.17 eV) has a much smaller adsorption energy than the $H_2O$ molecule (0.30 eV).

    To make a comparison, we also performed studies on the adsorption of $H_2O$ and $O_2$ molecules on the pristine $WSe_2$ monolayer. Different adsorption sites are considered for the adsorption of the molecules with various initial molecule orientations. The optimized structures for the most stable ones are shown in Fig. S(1) of ESI. It is clear that the Se vacancy can greatly enhance the adsorption strength of $H_2O$ and $N_2$ molecules. For $H_2O$ molecules, the adsorption energy on $V_{Se}$ is larger than that on the pristine $WSe_2$ monolayer by 67%, and the distances between the molecule and the basal plane of the pristine $WSe_2$ is larger than that of $V_{Se}$ by 1.7 Å. For $N_2$ molecules, the Se vacancy increases the adsorption strength by 0.05 eV.

    Previously, it is proposed that the physisorbed gas molecules can be detected sensitively through a charge transfer mechanism by the TMDs and other 2D materials. Therefore, we calculated the charge transfer between the adsorbed molecules and the $WSe_2$ monolayer by using Bader charge analysis. It is shown than $H_2O$ molecules gain electrons of 0.06 $e$ from $V_{Se}$, which is much larger than that from the pristine $WSe_2$ monolayer (0.02 $e$). The enhanced charge transfer is also observed for $N_2$ molecules. However, it is noted that the $H_2O$ molecules act as electron acceptors, while the $N_2$ molecules as electron donors. Apart from the enhanced charge transfer, the charge density differences (CDDs) are also studied. The CDD are calculated with respect to

the adsorbed molecule and the defective $WSe_2$ support with the configuration in the adsorbed system. Compared with results presented in Fig. 2 and Fig. S1, it is obvious that the Se vacancy can also enhance the charge redistribution due to the adsorption of the molecule. Overall, the $V_{Se}$ system exhibits a significantly enhanced sensing ability toward the $H_2O$ and $N_2$ molecules, as indicative of the enhanced adsorption energy, charge transfer, and charge redistribution. In addition, we also studied the electronic structure of the adsorbed systems, in terms of band structures and DOS (not shown). Our results reveal that the electronic states are almost unchanged upon the adsorption of molecules, for both the pristine and the defective $WSe_2$ monolayers.

### 3.3 Chemisorption of CO, NO, $NO_2$ and $N_2O$ molecules

### 3.3.1 CO and NO

Our extensive calculations show that CO, NO, $NO_2$ and $N_2O$ molecules are highly prone to be chemisorbed on the $V_{Se}$ surface. In this section, the adsorption behaviors of CO and NO are firstly presented and discussed. The adsorption energies, atomic structures, and relevant structural parameters for the PS, MC, and DC states of CO adsorption are shown in Fig. 3 (a). For the PS state, the adsorption energy of the CO is 0.23 eV. The C-O bond length (1.14 Å) of the adsorbed CO is almost unchanged compared with the free CO molecule. The distance between the C atom and the basal plane of $V_{Se}$ is 2.23 Å, which is accompanied by a large distance between the C atom and the atoms of the substrate (~ 4 Å). In addition, Bader charge analysis shows that the adsorbed CO is only slightly negatively charged by 0.03 $e$. The MC state for CO adsorption has an adsorption energy of 1.57 eV, which is much more exothermic than the PS state. The adsorbed CO is completely filled into the Se vacancy. The C-O bond length of the adsorbed CO is significantly elongated to 1.28 Å. The C atom also forms strong bonds with three original under-coordinated W atoms surrounding the Se vacancy. The lengths of these three C-W bonds are much smaller than those of the pristine $WSe_2$ (2.54 Å), among which two C-W bonds lengths are 2.13 Å and one C-W bond length is 2.24 Å. In addition, the C atom of the adsorbed CO lies below the upper Se atomic plane by 0.50 Å. These structural characteristics may result from the much smaller atomic radius of C than that of Se. It is noted that we also studied the MC state with the O atom occupying the Se vacancy. If is found that the filled CO molecule will run away from the vacancy after fully structural optimization. The DC state has the separated C atom occupying the Se vacancy and the O atom sitting above one neighboring Se atom (Se-O bond length 1.68 Å), which can be considered as the C-doped WSe2 monolayer with an adsorbed O atom on a Se atom neighboring the doped C. However, this state is highly endothermic compared with the MC state by 2.60 eV. Therefore, it is highly unfavorable for CO molecules to dissociate induced by the Se vacancy. And, from an energetic point of view, the CO molecule will exist in the MC state on $V_{Se}$.

The interaction of NO molecules with the $V_{Se}$ surface is much stronger than that of CO molecules. The adsorption energy of the NO for the PS state is 0.37 eV, larger than that of the CO molecule by 0.14 eV. The N-O bond length of the adsorbed NO for is elongated by 0.01 Å. The distance between the lower N atom of the adsorbed NO and the upper Se atomic plane is 1.30 Å. There is also more electron transfer (0.11 e) for the adsorbed NO than the adsorbed CO. The MC state for the NO adsorption is much more stable than that for the CO adsorption, with the adsorption energy difference up to ~ 1.6 eV between two states. The structural characteristics of the MC state for NO is similar to those of that for CO, which will be not further discussed. In addition, the NO molecule can adopt the configuration having the O atom occupying the Se vacancy, however, for which the adsorption energy is only 0.35 eV. Contrary to the CO molecule, the DC state for the NO adsorption is highly endothermic by 2.50 eV. This state can be considered as the N-doped WSe$_2$ monolayer with one O atom sitting above one Se atom neighboring the doped N. Although the DC state for NO is endothermic with respect to the free NO molecule, it is exothermic with respect to the MC state by ~ 0.7 eV. Further, our CI-NEB calculation (not shown) indicates that the transition from the MC to the DC states needs to overcome an energy barrier of 2.12 eV. Therefore, the NO molecule will also exist in the MC state on $V_{Se}$.

CDDs for the MC states of CO and NO molecules are also shown in Fig. 3 in order to gain further insight into the interaction between the molecules and the supports. As shown in Fig. 3(a), the transferred electrons are mainly localized on the C atom and the

C-W bonds, in agreement with the strong adsorption of CO. On the other hand, there is electron depletion on the C-O bond, and thus the C-O bond has been significantly weakened, which will make the O atom protruding above the S plane chemically active towards other reducing molecules, including CO itself. The CDD of the MC state for NO adsorption is shown in Fig. 3(b). which is similar to the case of CO adsorption.

Furthermore, the transition from the PS to the MC states for CO and NO molecules is discussed, and the PS state is selected as the initial state (IS) and the MC state as the final state (FS). From Fig. 4(a), it can be seen that, to fill the S vacancy, an energy barrier of 0.51 eV needs to be overcome, which is similar to the case of CO adsorption on the defective $MoS_2$ monolayer. It was proposed that an elementary reaction with an energy barrier less than 0.9 eV from the DFT calculation could occur at room temperature easily.[77] Therefore, the filling process of the Se vacancy by the CO molecule can even occur easily under conditions far below room temperature. The In the transition state (TS), the C-O bond length is slightly elongated by 0.01 Å compared with the IS, while the distance between CO and the upper Se plane of $V_{Se}$ is significantly reduced from 2.23 in the IS to 0.85 Å. For the NO molecule, the process of the filling of the Se vacancy is barrierless, as shown in Fig. 4(b), which is similar to the case of NO adsorption on the defective $MoS_2$ monolayer.[42, 43]

The electronic structures of the CO and NO doped $WSe_2$ monolayer are discussed. The band structures and the LDOS for the MC state of CO adsorption are shown in Figs. 5(a) and 5(b), respectively, while those for the MC state of NO adsorption are shown in Figs. 5(c) and 5(d), respectively. The band structures presented in Figs. 5(a) and 5(c) show that the $WSe_2$ are p-doped by substituting Se for the CO or NO molecule. This may be explained from the molecular orbital point of view. When CO acts as an electron acceptor, the transferred electrons will be accepted by the $2\pi^*$ anti-bonding orbitals (see Fig. S2(a)), which are doubly degenerate and can accommodate four electrons. For the NO molecule, the transferred electrons will also be accepted by the $2\pi^*$ anti-bonding orbitals (see Fig. S2(b)), however, which can only accommodate three electrons. Obviously, both the CO and NO molecules can accommodate more electrons than the Se atom, due to its $4s^24p^4$ electron configuration. Therefore, when the $WSe_2$ monolayer is doped by substituting Se atom for the CO or NO molecule, it will exhibit a p-type doping behavior. However, it is noted that the CO molecule may be not appropriate for doping the WSe$_2$, as the defect level still belongs to the deep acceptor level.

The LDOS of the C and O atoms, and the three W atoms marked in Fig. 3(a) are shown in Fig. 5(c). It can be seen that the defect level below the Fermi level (E$_f$) mainly comes from the O and C 2p states (the 2π* orbitals of the CO), and the W1 and W2 5d states, while the W3 5d states make an important contribution to the defect level above the E$_f$, except the 2π* orbitals of the CO. The MC state for the NO adsorption is magnetic with a spin magnetic moment of 1.00 $\mu_B$. There is one defect level for the spin-up state and two defect levels for the spin-down states, which also mainly comes from the 2π* orbitals of the adsorbed molecule and the 5d states of the W atoms bonded with the N atom. Finally, Bader charge analysis shows that the adsorbed CO and NO molecules have gained electrons of 2.22 and 1.49 $e$, respectively. This is in agreement with the fact that upon adsorption considerable empty 2π* orbitals of the both molecules shift down below the E$_f$.

It is meaningful and necessary to see if the adsorbed CO (NO) molecule in the MS state can react with further injected molecules. For such purpose, the interaction of the second CO (NO) molecule with the preadsorbed CO (NO) molecule on the V$_{Se}$ surface is studied. For the CO molecule, it can be physisorbed above the preadsorbed CO molecule (the IS in Fig. 6), and it can also combine with the O atom of the preadsobred CO to form a physisorbed CO$_2$ above the C doped WSe$_2$ monolayer (the FS in Fig. 6). Our CI-NEB calculation shows that the transition between two states needs to overcome an energy barrier of 0.89 eV. For the FS, the calculated adsorption energy is 0.29 eV, with respect to the free CO$_2$ molecule. Therefore, it is very easy for the formed CO$_2$ to desorb from the C doped WSe$_2$, leaving a C doped WSe$_2$ monolayer. For NO, it is found that even if initially the molecule is away from the surface by a distance up to 2.9 Å (Fig. S3(a)), after optimization it still can directly combine with the O atom of the preadsobred NO molecule to produce a physisorbed NO$_2$ molecule above the N doped WSe$_2$ monolayer (Fig. S3(a)). Energetically, the state in Fig. S3(b) is more stable by 1.26 eV than the state shown in Fig. S3(a). And the state in Fig. S3(b) has an adsorption energy of 0.53 eV with respect to the free NO$_2$ molecule. Therefore, the final product for NO doping of V$_{Se}$ is probably the N doped WSe$_2$.

The atomic and electronic structures of the C-doped WSe$_2$ and the N-doped WSe$_2$ investigated. Both the C-doped and the N-doped WSe$_2$ are nonmagnetic. From Fig. 7(a), it can be seen that the doped C (N) atom lies below the upper Se plane, forming three bonds with the surrounding W atoms with the bond lengths about 2 Å. From Fig. 7(b), it is clear that both the C-doped and the N-doped WSe$_2$ monolayer are p-doped, as expected. The defect levels for both cases mainly come from the p states of the dopants and the d states of its neighboring W atoms. Similar to the molecule doping, the N doping is also better than the C doping.

In addition, according to the above result, the significant experimental result of Ref.[53] may be reinterpreted. In this work, Zhao et al. explored the p-doping of WSe$_2$ via chemisorption of NO$_x$ at a sample temperature of 150 °C. The DFT calculations and spectroscopy results identified the chemisorbed NO as the dominant doping species at the Se vacancies sites with the N atom occupying the vacancy.[53] However, according to our theoretical calculation, the exposed O atom of the filled NO in the Se vacancy can be removed, even spontaneously, by another NO molecule. This will produce a N-doped WSe$_2$, which also exhibits n-type doping behavior. Therefore, it is suggested that the air stable p-doing found in Ref.[53] may simply be due to the replacing of the Se atom with the N atom, besides the filled NO molecule.

**3.3.2 NO$_2$ and N$_2$O**

The results on the adsorption of the NO$_2$ and N$_2$O molecules on the V$_{Se}$ are presented and discussed in this section. The PS states for the NO$_2$ and N$_2$O molecules are shown in Figs. 8(a) and 8(c), respectively. The PS state for the NO$_2$ molecule has an adsorption energy of 0.36 eV. The distance between the NO$_2$ molecule and the substrate is 1.56 Å and the nearest distance between the molecule and the support is close to 4 Å. However, the interaction leads to a change of the molecule structure. The bond lengths of both N-O bonds are slightly elongated (by 0.01 or 0.02 Å), while the bond angle is reduced by ~ 5º, which may be due to that the electron transfer from the support to the molecule (0.20 $e$). The interaction strength of the N$_2$O molecule and V$_{Se}$ in the PS state is weaker than that of $N_2O$. The PS state for the $N_2O$ molecule has an adsorption energy of 0.24 eV. Accordingly, the distance between the $N_2O$ and the upper Se plane (2.08 Å) is also large compared with the case of $N_2O$ adsorption. And there is a minor electron transfer (0.03 $e$) to the adsorbed $N_2O$ molecule.

For the chemisorption, it is found that there are only DC states for both the $NO_2$ and $N_2O$ molecules. As shown in Fig. 8(b), the adsorbed $NO_2$ can completely dissociate to form a O-doped $WSe_2$ with an adsorbed NO molecule. The dissociation process is endothermic by 3.82 eV with respect to the free NO2 molecule. The distance between the adsorbed NO and the doped O atom is 2.65 Å, which can exclude the possibility of NO forming chemical bond with the support. The bond length (1.17 Å) of the adsorbed NO is hardly unchanged compared with the free molecule, further confirming its weak adsorption. The adsorption energy for the NO molecule is also studied, with respect to the free NO molecule and the O-doped WSe2 monolayer. An adsorption energy of 0.32 eV suggests that the physiorbed NO can be easily desorbed from the support, leaving a O-doped WSe2 monolayer. For the DC state of the $N_2O$ molecule, as shown in Fig. 8(d), the dissociation of the $N_2O$ molecule on $V_{Se}$ will produce a O-doped $WSe_2$ with an physisorbed N2 molecule. The N2O dissociation process is also highly endothermic (by 5.06 eV) with reference to the free molecule. The distance between the physisorbed $N_2$ and the doped O atom is 3.06 Å, and the adsorption energy for the $N_2$ is 0.20 eV. These suggest that the upon N2O dissociation on the $V_{Se}$, there will be a O-doped $WSe_2$ monolayer, similar to the case of the $NO_2$ adsorption. In addition, the possible chemisorption state with the N atom of the N2O pointing down has also been studied (Fig. S4). It is found that this state is much less stable than the state shown in Fig. 8(d) by about 2.7 eV, thus, which will be not further investigated.

The processes of the transition from the PS state to the DC state are shown in Fig. 9(a) and 9(b), respectively, for the $NO_2$ and $N_2O$ molecules. For the $NO_2$ molecule (Fig. 9(a)), the dissociation process is highly endothermic by 3.46 eV and only needs to overcome an energy barrier of 0.04 eV. For the TS, the adsorbed $NO_2$ molecule is much closer to the support than the IS, while the lower N-O bond is only elongated by 0.04 Å compared with the IS. For the $N_2O$ molecule, the dissociation process is much energetically favorable than that for the $NO_2$ molecule. However, the $N_2O$ molecule needs to overcome a large energy barrier of 0.98 eV. This may be due to the geometric structure of the TS for the N2O molecule. As shown in Fig. 9(b), in the TS, the $N_2O$ molecule is significantly distorted compared with that in the IS. The N-O bond is elongated by 0.12 Å and the molecule becomes bent having a bond angle of 149º.

The geometric and electronic structures of the O-doped $WSe_2$ are shown in Fig. 10. As shown in Fig. 10(a), the doped O atom lies about 0.70 Å below the upper S plane, and forming three O-W bonds of 2.08 Å. The O-doped $WSe_2$ monolayer is nonmagnetic. The band structure shown in Fig. 10(b) shows that the defect levels induced by the Se vacancy are completely removed, which is in good agreement with previous results. The LDOS projected on the doped O and its bonded W show a significant hybridization between the O 2p states and the W 5d states, revealing a strong interaction between two species. Previously, a surface laser modification was proposed to passivate the Se vacancy by O atoms, by which the conductivity and the photoconductivity are enhanced by 400 and 150 times compared with the as-grown $WSe_2$. Our studies suggest that passivating Se vacancy by O atoms may be realized through a chemical route, by first dosing $NO_2$ or $N_2O$ gas onto the $WSe_2$ surface and then heating the surface to desorb the physisorbed NO or $N_2$. It is noted that the produced $N_2$ is safe, unlike NO, which is toxic and harmful. Therefore, although for $N_2O$ the process needs to operate at an elevated temperature compared with the room temperature, due to its large energy barrier of ~ 1 eV, it is proposed that the $N_2O$ gas may be more favorable for healing the Se vacancy than $NO_2$.

## 4. Conclusions

In this study, we theoretically investigated the effect of Se vacancies on the interaction between the gas molecules, including $H_2O$, $N_2$, CO, NO, $NO_2$ and $N_2O$, with the $WSe_2$ monolayer surfaces. According to the calculated energies, $H_2O$ and $N_2$ molecules are highly prone to be physisorbed on the $V_{Se}$ surface. Compared with the perfect lattice site, the presence of the Se vacancy can significantly enhance the sensing ability of the $WSe_2$ monolayer toward the $H_2O$ and $N_2$ molecules via a surface charge transfer mechanism, as indicative of the enhanced adsorption energy, charge transfer, and charge redistribution. In contrast, CO and NO molecules highly prefer to be molecularly chemisorbed on the $V_{Se}$ surface with the non-oxygen atom occupying the Se site. Further, the exposed O atoms of the molecularly chemisorbed CO or NO can be reacted with the additional CO or NO molecule, to produce the C-doped or N-doped $WSe_2$ monolayer. Our electronic structure calculations show that the $WSe_2$ monolayers are p-doped by the CO and NO molecules, as well as the C and N atoms. The calculated energies suggest that the filling of the CO or NO molecule and the removal of the exposed O atom by the additional CO or NO molecule are both energetically and dynamically favorable, which at least can occur easily at room temperature. However, only the NO molecule and the N atom doped $WSe_2$ monolayers exhibit significantly improved electronic structures compared with the defective $WSe_2$ monolayer. For the $NO_2$ and $N_2O$ molecules, they are highly prone to dissociate directly to form a O-doped $WSe_2$ with a physisorbed NO or $N_2$. After desorbing the physisorbed NO or N2, the O-doped $WSe_2$ monolayer can be obtained, for which the defect levels due to the Se vacancy can be completely removed. The calculated energies suggest that although the dissociation processes for $NO_2$ and $N_2O$ molecules are highly endothermic, the $N_2O$ dissociation may need to operate at an elevated temperature compared with the room temperature, due to its large energy barrier of ~ 1 eV. The present study will give novel insight into the doping of $WSe_2$ 2D semiconductors via the interaction between common gas molecules and the Se vacancy, and also reveal the possible application of the defective $WSe_2$ in the gas sensing field.

**Acknowledgements:**

This work is supported by the Henan Joint Funds of the National Natural Science Foundation of China (Grant No. U1504108), National Natural Science Foundation of China (Grant Nos. 11674083, 11447001, and 21603109). Innovation Scientists and Technicians Troop Construction Projects of Henan Province (No. C20150029).

**Table 1.** Calculated adsorption energies (in eV) for $H_2O$, $N_2$, CO, NO, $NO_2$, and $N_2O$ at various adsorption states on $V_{Se}$.

| molecule | $H_2O$ | $N_2$ | CO | NO | $NO_2$ | $N_2O$ |
|---|---|---|---|---|---|---|
| PS | 0.30 | 0.17 | 0.23 | 0.37 | 0.36 | 0.24 |
| MC | --- | -0.08 | 1.57 | 3.15 | --- | --- |
| DC | --- | --- | -1.03 | 2.50 | 3.82 | 5.06 |

**Figure Captions:**

**Fig. 1.** (a) Top and side views of the VSe system, which is created by removing a single Se atom from the (4×4) $WSe_2$ monolayer. The red dashed circle indicates the position of the Se vacancy. The light green and grey spheres represent the Se and W atoms, respectively. (b) The band structure of the $V_{Se}$ system. The horizontal dashed blue line indicates the $E_f$. (c) The spin-polarized TDOS of the $V_{Se}$ system and the LDOS projected on the 5d states of the W atoms neighboring the Se vacancy. The positive and negative DOS denote the spin-up and spin-down states, respectively. The vertical dashed line indicates the $E_f$.

**Fig. 2.** The top and side views of the $V_{Se}$ with the physisorbed $H_2O$ (a) and $N_2$ (b) molecules. The adsorption energy (in eV), the number the transferred electrons (in $e$), and the height (h in Å) of the adsorbed molecule with respect to the upper Se atomic plane are given. The isosurfaces for the CDD are $2\times10^{-4}$ and $1\times10^{-4}$ $e$ bohr$^{-3}$, respectively, for the adsorption of the $H_2O$ and $N_2$ molecules. The red and green regions represent the electron accumulation and depletion, respectively.

**Fig. 3.** The top and side views of the PS, MC and DC states for CO and NO molecules adsorption on the $V_{Se}$ systems. The adsorption energy (in eV), the bond length of the adsorbed molecule, and the height (h in Å) of the adsorbed molecule with respect to the upper Se atomic plane are given. The distances between the C or N atom and its neighboring W atoms are also displayed for the MC and MC states. The isosurface for the CDD of the MC state is taken as $5\times10^{-3}$ $e$ bohr$^{-3}$. The red and green regions represent the electron accumulation and depletion, respectively.

**Fig. 4.** (a) Atomic configurations of the IS, TS, and FS along the minimum-energy path for the transition from the PS to MC states for the CO adsorption. (b) Variation of the energy for the transition from the PS to MC states for the NO adsorption. In (a) and (b), the energies are given with respect to the total energy of the PS states.

**Fig. 5.** The band structures and the relevant LDOS are shown in (a) and (b), respectively, for the CO adsorption, while those for NO adsorption in (c) and (d), respectively. The W1, W2, and W3 atoms are marked in Fig. 3, which are bonded with the C or N atom of the adsorbed CO or NO molecule. It noted that for the CO adsorption, the MC state is nonmagnetic, and only the spin-up states are shown in (a) and (b).

**Fig. 6.** Atomic configurations of the IS, TS, and FS along the minimum-energy path for the removal process of the exposed O atom for the CO adsorption. The relevant structural parameters (in Å) and the energies of the IS, TS, and FS are given, which are respect to the total energy of the IS.

**Fig. 7.** (a) Top and side views of the C and N-doped WSe$_2$ monolayers. The band structures of the C and N-doped WSe$_2$ monolayers are shown in (b) and (c), respectively. The LDOS projected on the doped C or N atom and its bonded W atoms are shown in (d). It noted that both systems are nonmagnetic, and only the spin-up states are shown. In addition, in the lower panel of (d), a inset showing the DOS near the E$_f$ is given.

**Fig. 8.** The PS and DC states for the NO$_2$ molecule adsorption are shown in (a) and (b), respectively, while those for the N$_2$O molecule adsorption shown in (c) and (d).

**Fig. 9.** Atomic configurations and energies of the IS, TS, and FS along the minimum-energy path for the transition from the PS to DC states for the NO$_2$ molecule adsorption (a) and the N$_2$O molecule adsorption (b).

**Fig. 10.** (a) Top and side views of the O-doped WSe$_2$ monolayers, for which the band

structure is shown in (b). In (c), the LDOS are those projected on the doped O and its bonded W atoms. The system is nonmagnetic, and only the spin-up states are shown.

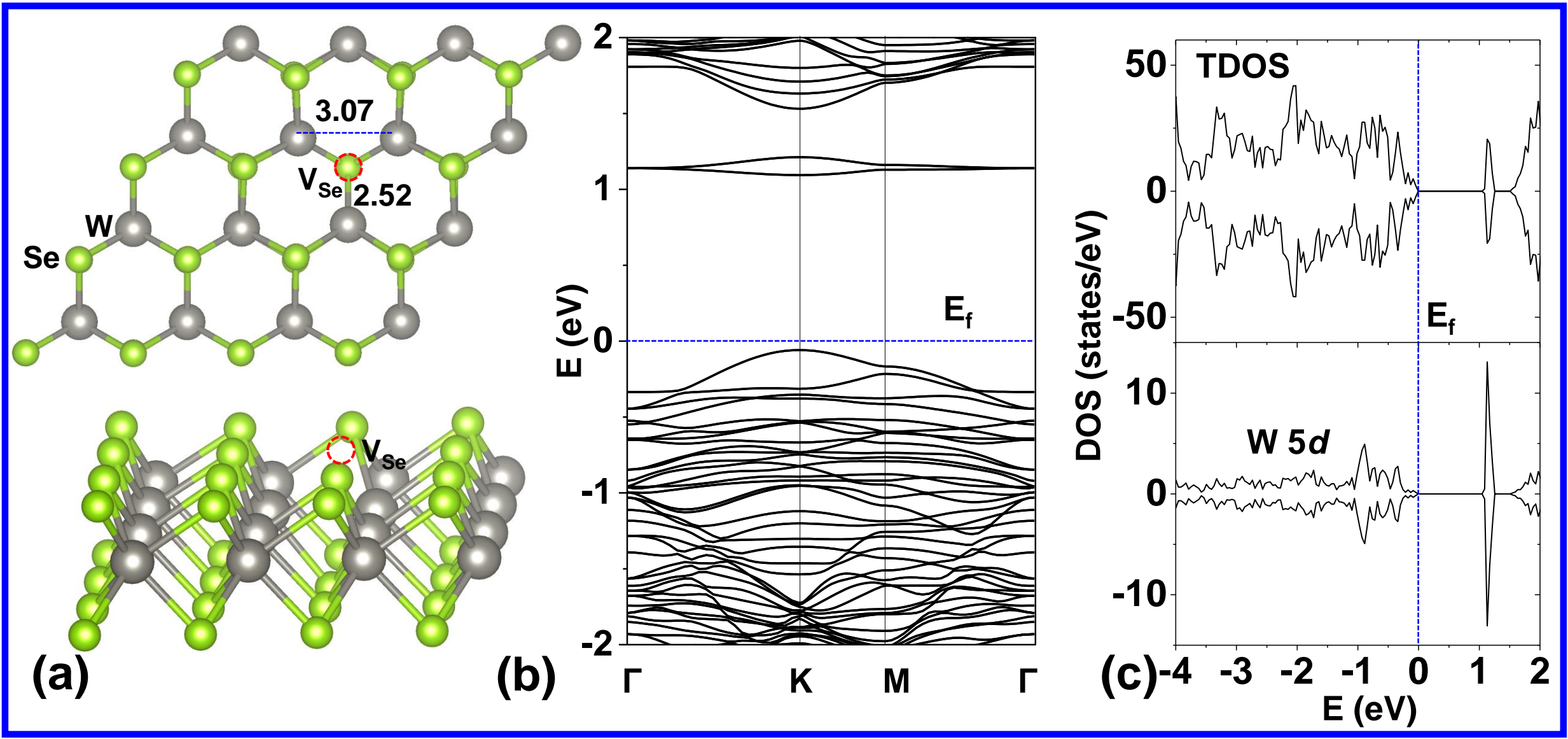

Figure 1

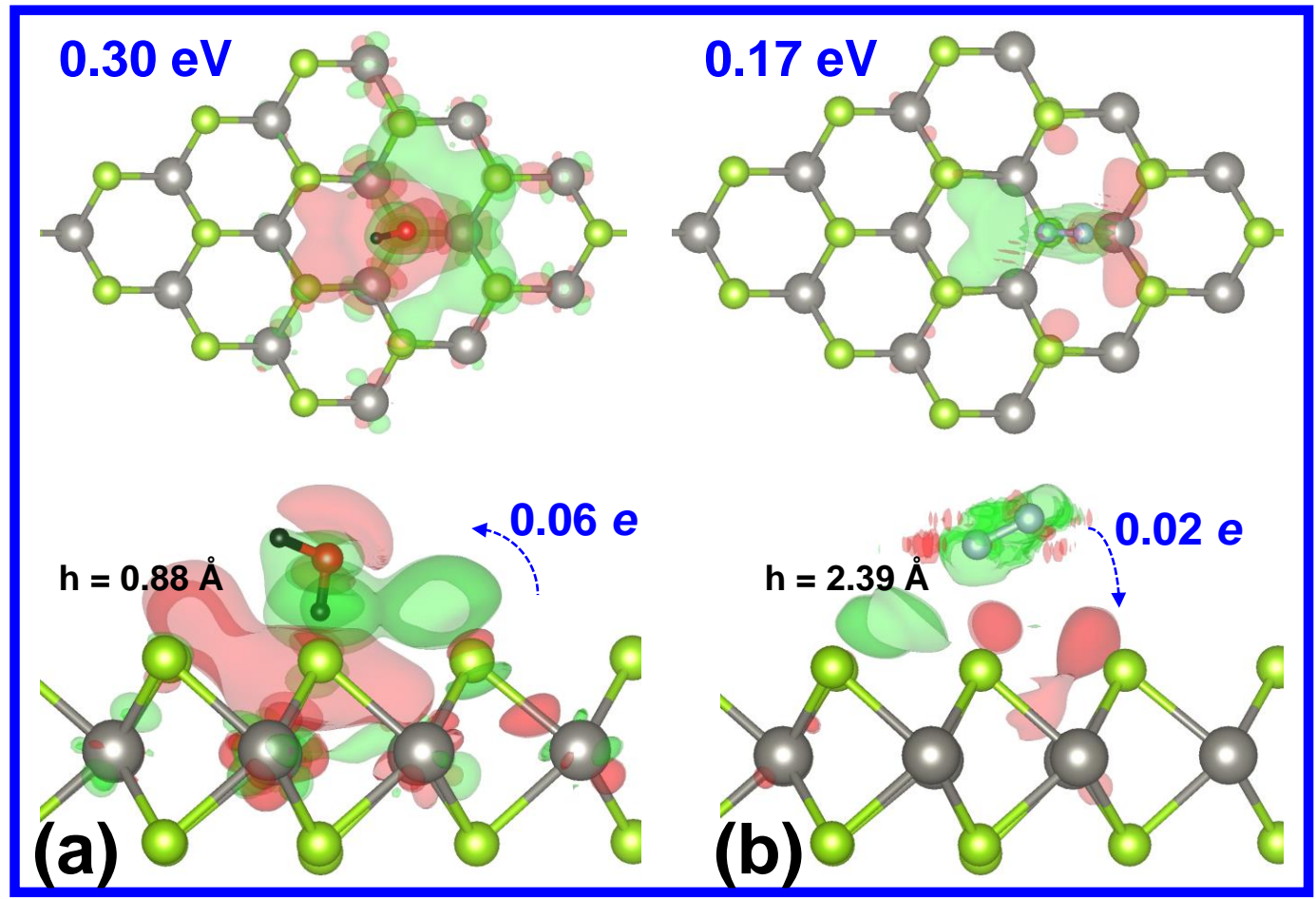

Figure 2

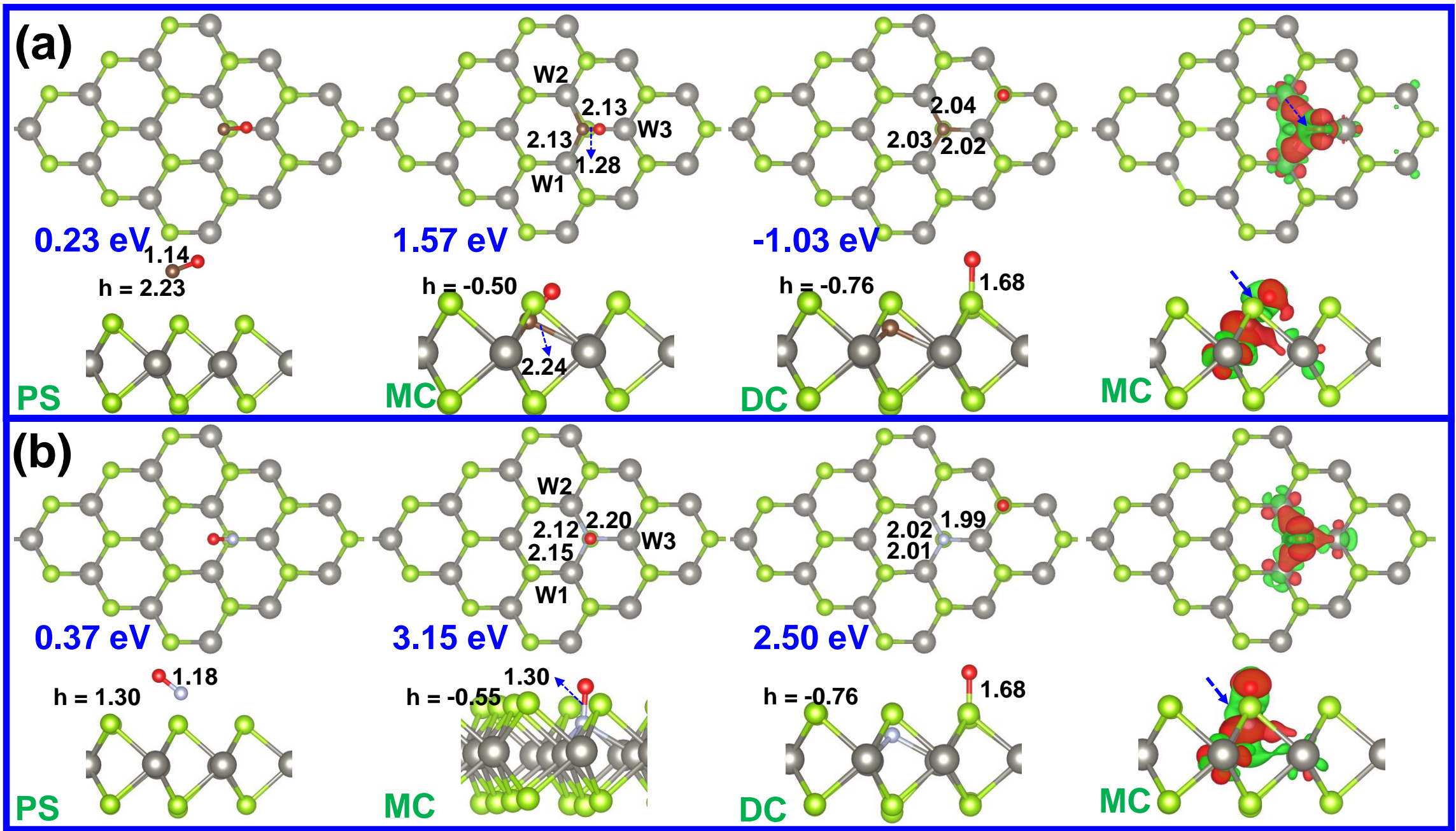

Figure 3

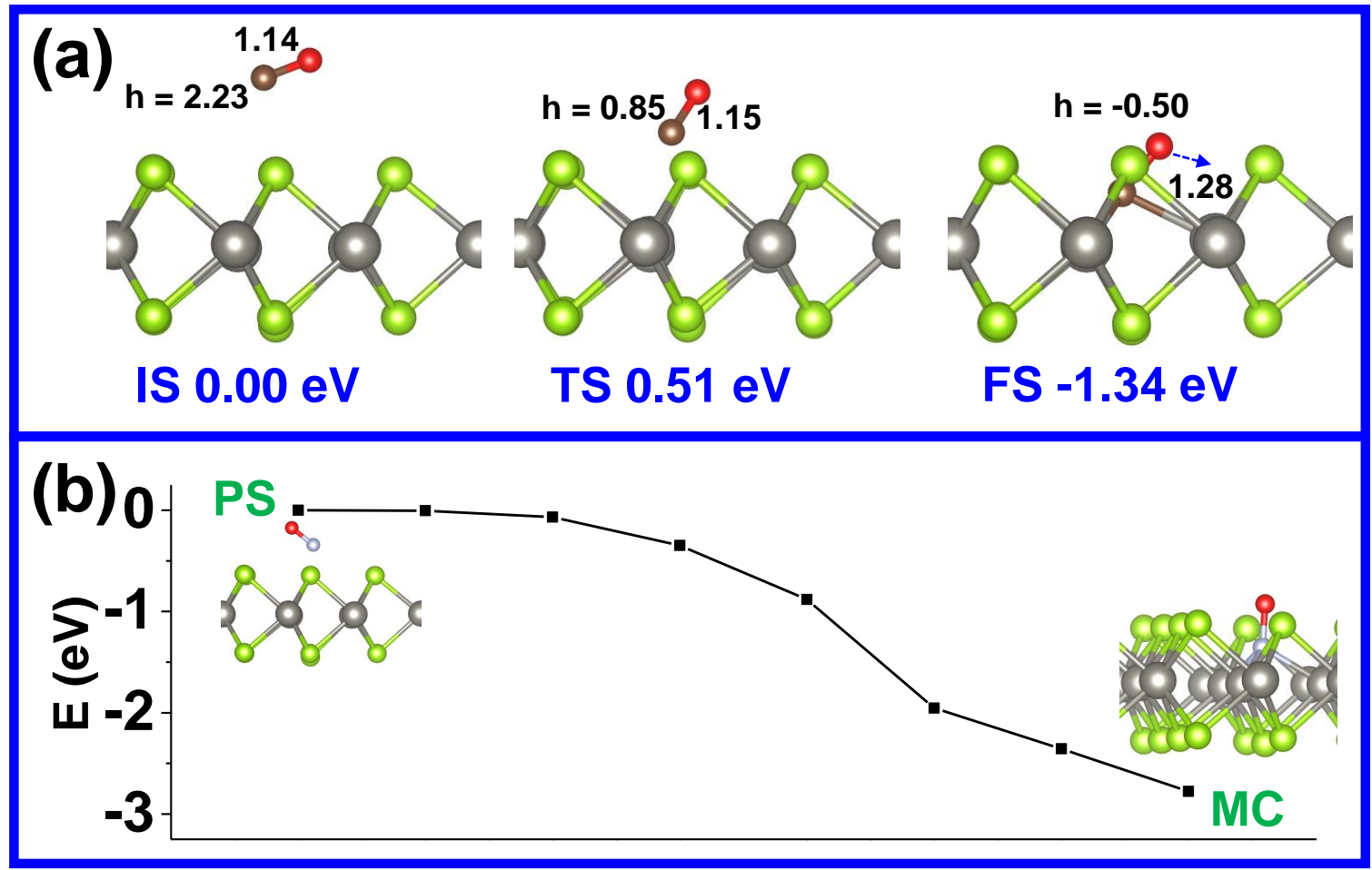

Figure 4

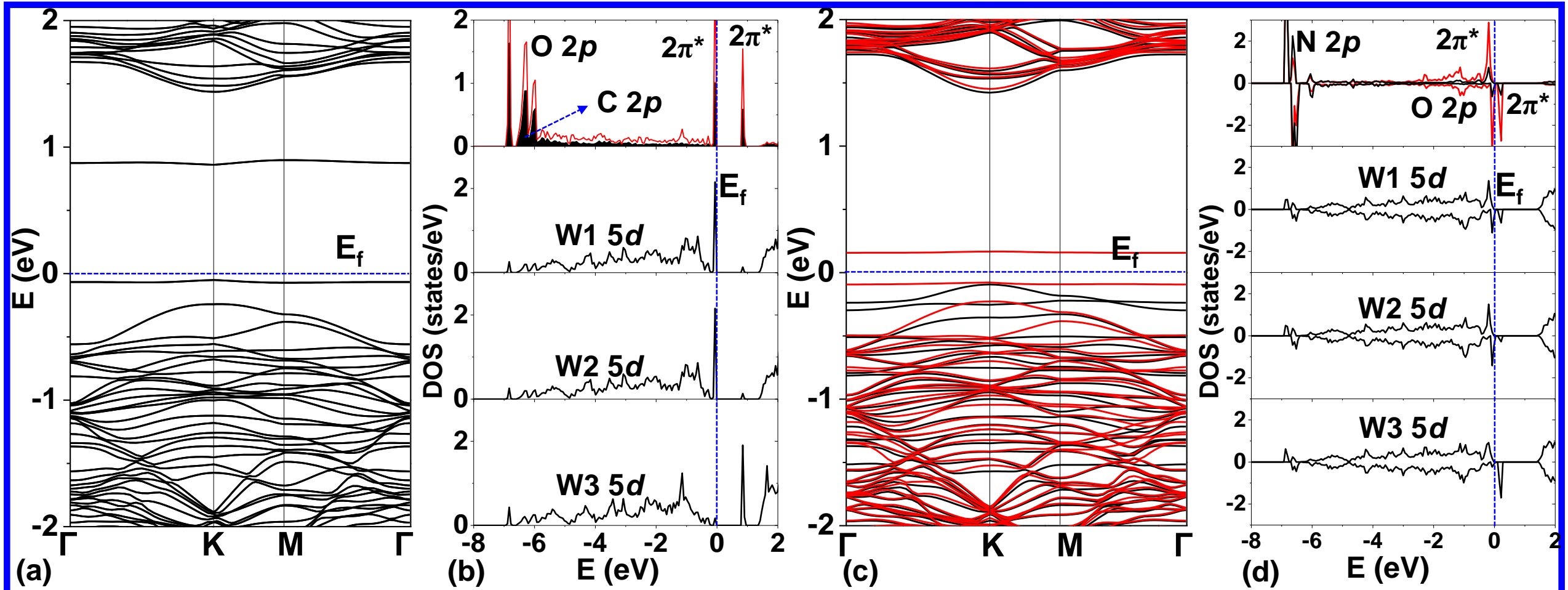

Figure 5

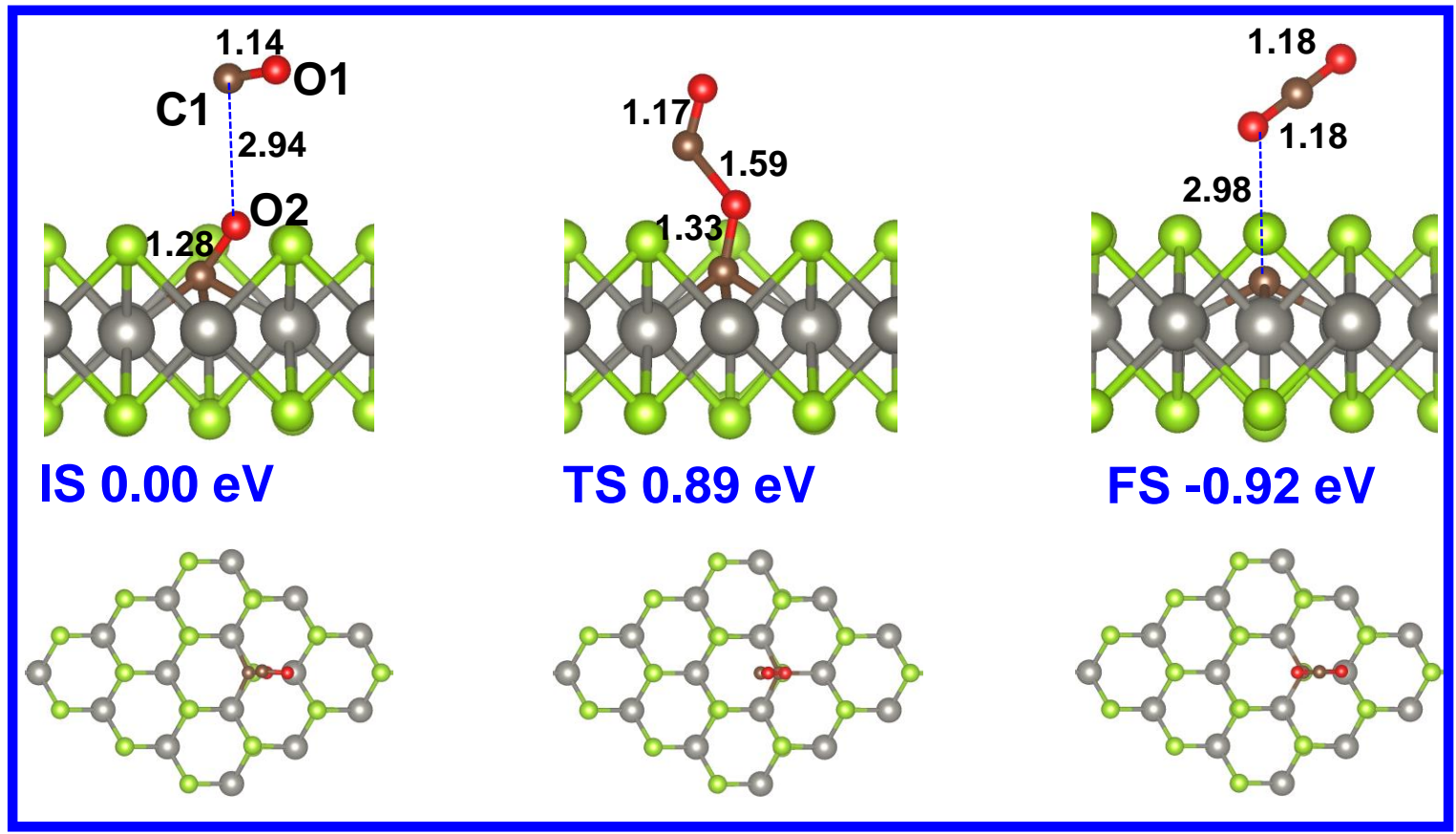

Figure 6

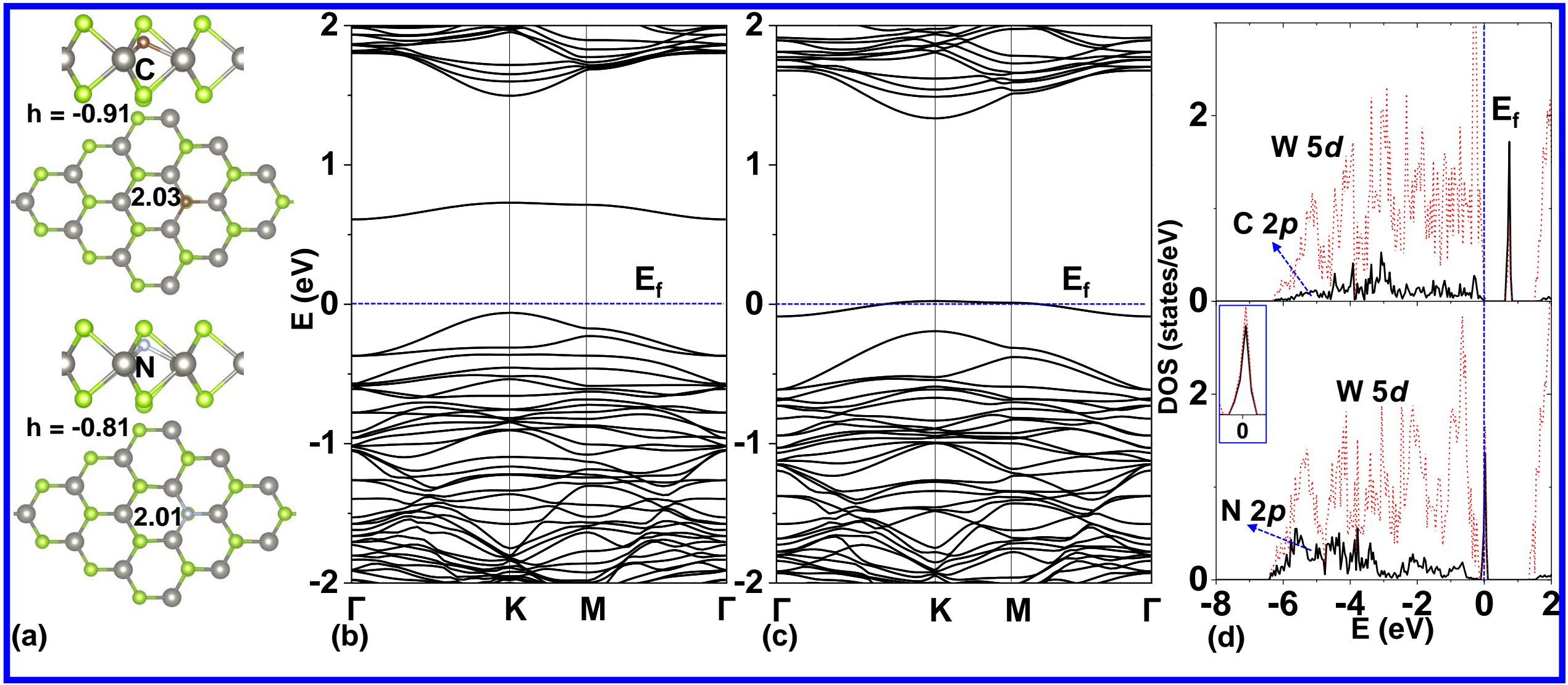

Figure 7

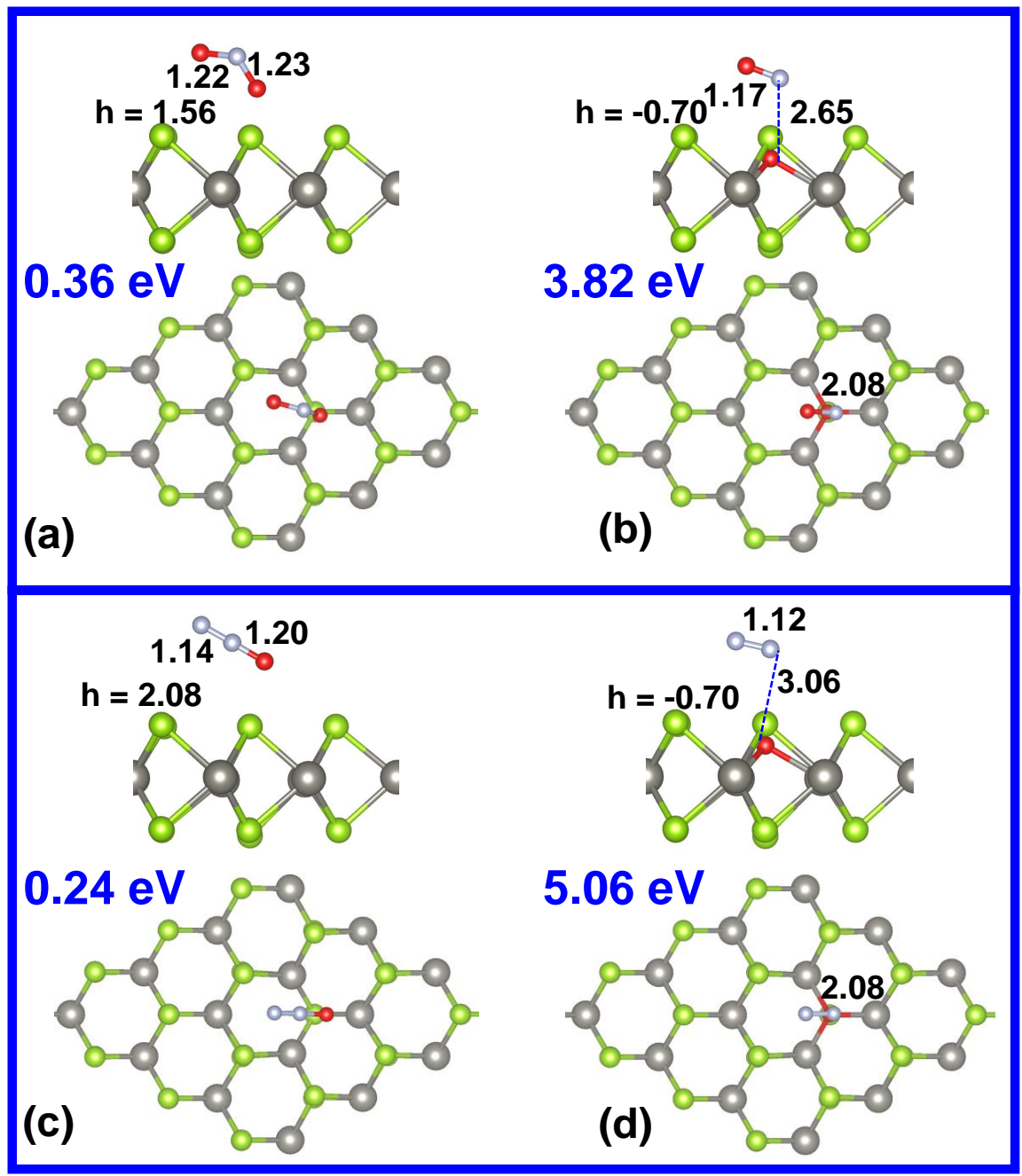

Figure 8

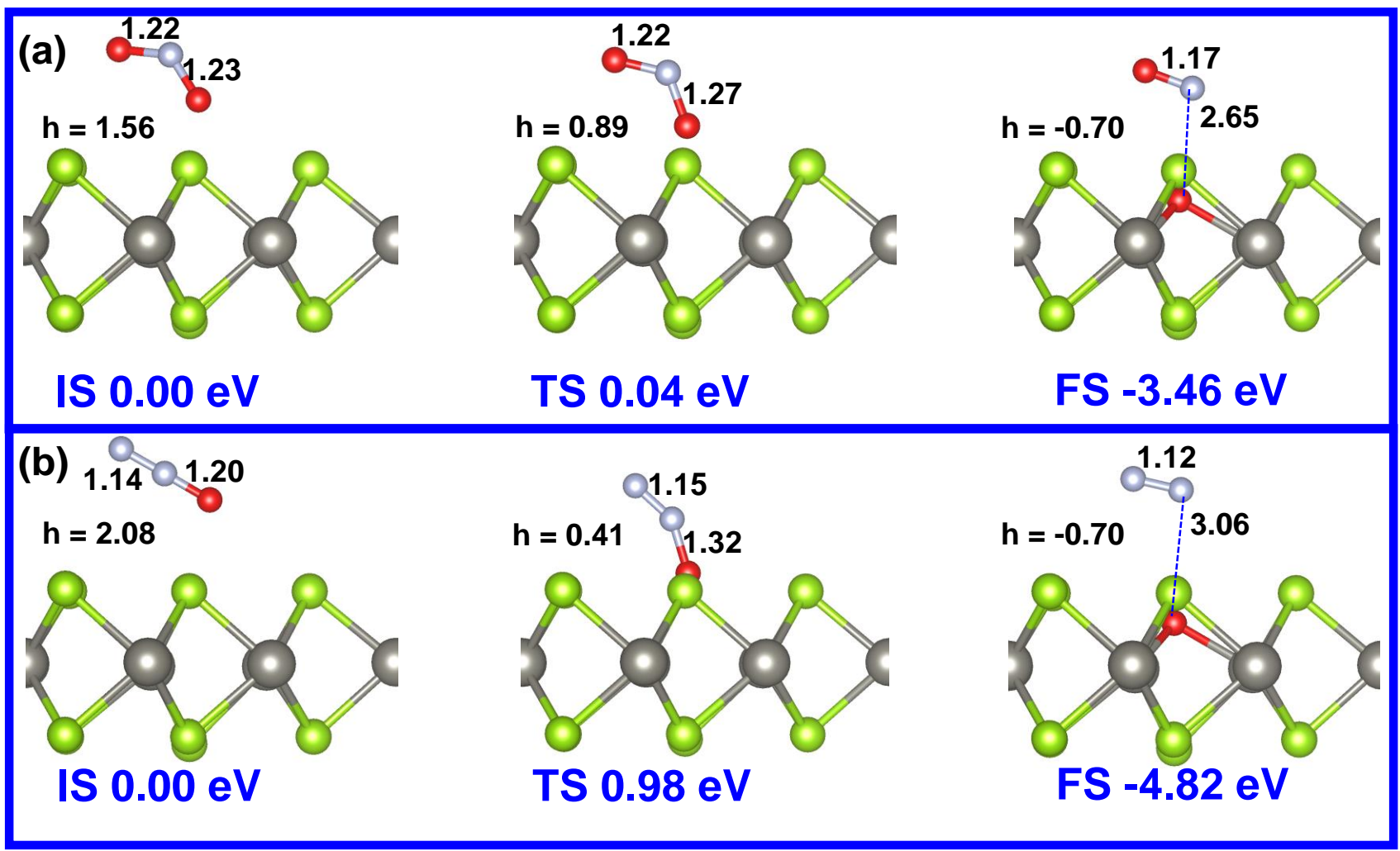

**Figure 9**

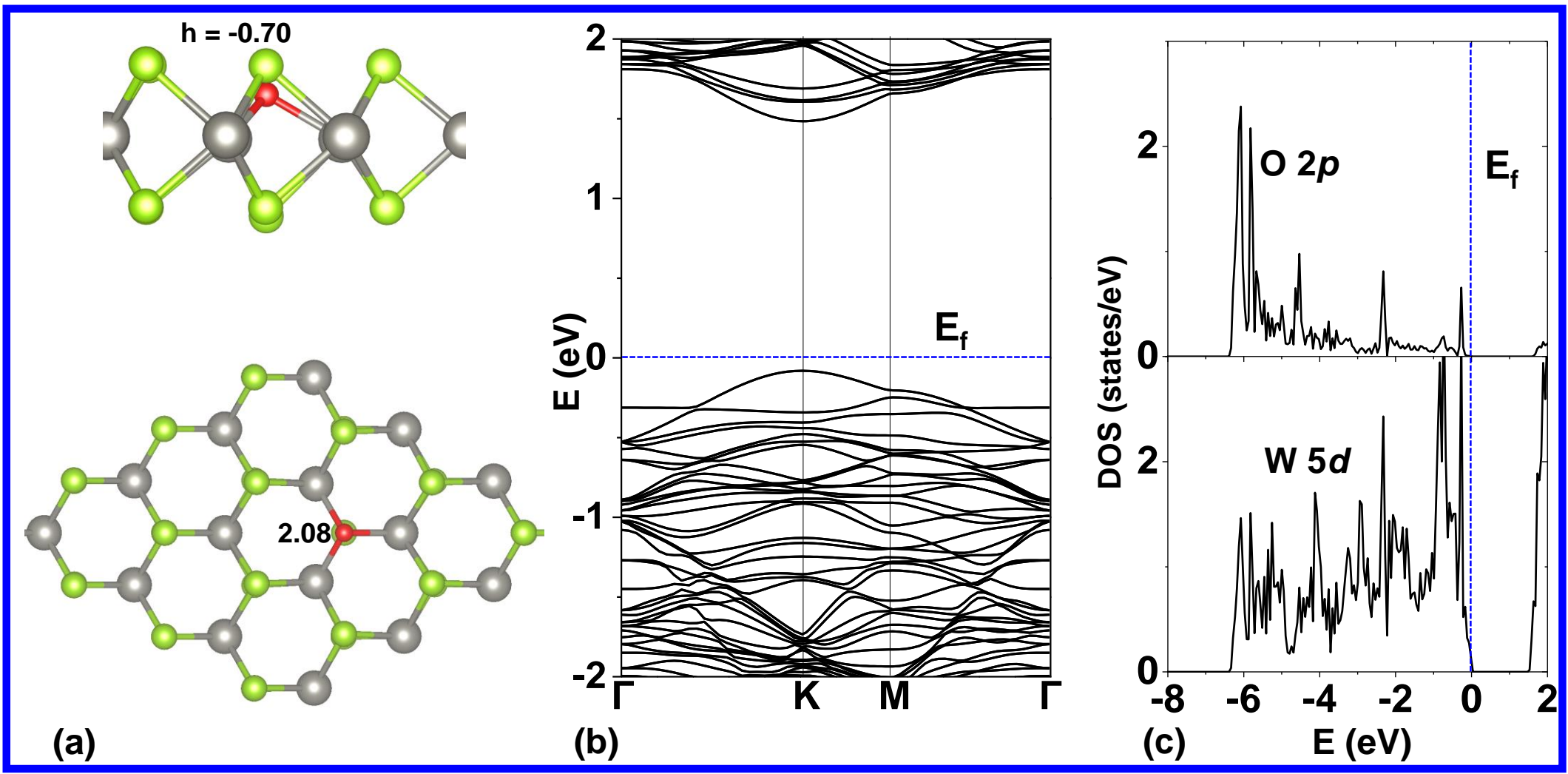

Figure 10